\documentclass[twocolumn,amssymb, amsmath,nobibnotes,nofootinbib,showpacs, aps, prb,a4paper]{revtex4-1}

\usepackage[dvips]{graphicx,color}

\begin{document}

\title{Natural ferroelectric order near ambient temperature in HoFeO$_3$: A member of $R$FeO$_3$ orthoferrites}
\author{K. Dey,$^{1,2}$ A. Indra,$^{1,3}$ S. Mukherjee,$^1$ S. Majumdar,$^1$ J. Strempfer,$^{4}$ O. Fabelo,$^{5}$ E. Mossou,$^5$ T. Chatterji,$^5$ and S. Giri$^1$}
\affiliation{$^1$School of Physical Sciences, Indian Association for the Cultivation of Science, Jadavpur, Kolkata 700032, India}
\affiliation{$^2$Department of Physics, SBSS Mahavidyalaya, Goaltore, Paschim Medinipur, W. B. 721128, India}
\affiliation{$^3$Department of Physics, Srikrishna College, Bagula, Nadia, W. B. 741502, India}
\affiliation{$^4$Deutsches Elektronen-Synchrotron DESY, Notkestrasse 85, D-22607 Hamburg, Germany}
\affiliation{$^5$Institut Laue-Langevin, 71 Avenue des Martyrs, CS 20156, 38042 Grenoble Cedex 9, France}

\begin{abstract}
 
Current scenario in multiferroics demands a breakthrough discovery of promising materials after BiFeO$_3$. Recently, the controversial discovery of room temperature ferroelectricity (FE) in SmFeO$_3$ [PRL {\bf 107}, 117201 (2011); {\bf 113}, 217203 (2014)] inspires the investigation of HoFeO$_3$. Here, we report a natural ferroelectric order below 210 K ($T_{FE}$) along $c$-axis with reasonably large polarization and low-field strong magnetoelectric coupling. Synchrotron and neutron diffraction results confirm that a shift of O atoms along $c$-axis of polar $Pb^{\prime}n^{\prime}2_1$ structure causes FE in HoFeO$_3$. The exchange striction mechanism is suggested to elucidate the ferroelectric order. The results create a renewed attention for searching promising candidates with a natural ferroelectric order and higher $T_{FE}$ in the rest of the $R$FeO$_3$ series.
    
\end{abstract}
\maketitle

Spin dependent electronics is identified to be the next-generation technology, that prevails over current generation of electronics solely based on the charge degrees of freedom. In particular, the control of magnetism by applying the external electric field is a great challenge in materials science for developing low power spintronic devices\cite{spaldin,fie,cheo,scott}. Multiferroics, belonging to a chemically single phase rare class of materials, are promising to meet the above criterion. The magnetoelectric (ME) coupling, as a result of cross coupling between magnetic and ferroelectric (FE) orders, leads to the tunability of magnetic polarization by the applied electric fields ($E$). In order to explore it technologically, this functionality should be observed close to the room temperature. Therefore, searching for the new promising materials after BiFeO$_3$ is one of the prime focuses in multiferroics, where the large FE polarization ($P$) associated with a reasonable ME coupling is desirable close to the room temperature\cite{wang}. The cupric oxide, CuO (tenorite), is an another promising inorganic material, that exhibits a spontaneous polar order, however, in a limited temperature region of 213$-$230 K\cite{kimura_NM,wang_nat_commun}. 

Recently, few members of the rare earth ($R$) ferrite ($R$FeO$_3$) has been found to hold the great promise of room temperature multiferroic order in the artificial hexagonal structures\cite{jeong_JACS,mukherjee_PRL,jeong,cheng}. 
The ferroelectricity at room temperature is observed for epitaxially grown YbFeO$_3$\cite{jeong_JACS}, GaFeO$_3$\cite{mukherjee_PRL} and LuFeO$_3$\cite{jeong} thin films, where a hexagonal heterostructure has been adopted in all the cases. More recently, the interface strain-induced ferroelectricity has also been proposed for SmFeO$_3$ films, which was epitaxially grown on a [001] SrTiO$_3$:Nb substrate with a proposed rhombohedral or hexagonal structure\cite{cheng}. The compound SmFeO$_3$ has been the center of attention, when the spin-canting driven natural ferroelectricity was reported above room temperature along the crystallographic $b$-axis\cite{lee_PRL}. On the contrary, a $G$-type collinear antiferromagnetic (AFM) structure has been established and suggested for the non-existent  ferroelectricity in SmFeO$_3$\cite{kuo_PRL}. More recent results on  polycrystalline SmFeO$_3$ again reported an evidence of the  spontaneous electric polarization at 173 K, as confirmed by the  polarization loop ($P-E$) measurements\cite{zhang_JALCOM}. In fact, the {\it first principles} calculations proposed a possible occurrence of electric polarization along $b$-axis associated with the $Pmc2_1$ ground state, where the exchange striction mechanism involved a polar displacement of the oxygen ions correlated with the proposed ferroelectricity.\cite{yang_cal} These contradictory results create arguments on the occurrence of intrinsic ferroelectricity in SmFeO$_3$. Unlike above mentioned $R$FeO$_3$, the AFM GdFeO$_3$ showed a FE transition along the crystallographic $c$-axis at the Gd ordering temperature of 2.5 K\cite{toku_NM,zhao_PRB}, that was further supported by the {\it ab initio} calculations considering an exchange striction between adjacent Fe and Gd atoms\cite{strop_NJP}. Below the Dy ordering transition at 3.5 K, the ferroelectricity was also observed for DyFeO$_3$ with a magnetic field ($H$) applied along the $c$-axis\cite{toku_PRL}. Importantly, the application of magnetic field causes a change of Fe spin configuration to the original high temperature $G_{x}F_{z}$ structure from $A_{x}G_{y}$ and points to the fact that the ferroelectric order in DyFeO$_3$ appears in the high temperature spin structure. 

The compound of our interest, the orthoferrite HoFeO$_3$, exhibits a long range magnetic order governed by the Fe$-$Fe, $R-$Fe, and $R-R$ interactions. The Fe spins of the compound order antiferromagnetically ($T_N$) around 647 K, which is followed by two more transitions driven by the $R-$Fe and $R-R$ interactions with decreasing temperature\cite{koe_PR,bujko_JPCM,gram_ZF,mare_ZP,chat_AIP,chat_arxiv}. As a result of the $R-$Fe interaction, the Fe spins reorient along the $c$-axis below $\sim$ 51 K and transform to a $G_{z}F_{x}$ phase from $G_{x}F_{z}$ in Bertaut's notation. The Ho order was further proposed to happen around $\sim$ 3.3 K from the heat capacity,  magnetization results, and M\"{o}ssbauer studies for HoFeO$_3$\cite{bhatt_JPCS,shao_JCG,nikol_JPCM}. 

In this letter, we report the onset of polar order along the crystallographic $c$-axis below $\sim$ 240 K for HoFeO$_3$ with a FE transition ($T_{FE}$) at $\sim$ 210 K. Importantly, the $P$ appears in the magnetic $G_{x}F_{z}$ state. This is exactly the same magnetic $G_{x}F_{z}$ state of DyFeO$_3$ with non-zero $H$\cite{toku_PRL} and GdFeO$_3$ with $H$=0\cite{toku_NM}, where ferroelectricity was observed at very low temperatures below the respective rare earth ordering temperatures. Here, the value of $P$ is significantly higher as $\sim$ 0.29 $\mu$C/cm$^2$ for a 10 kV/cm poling field. A considerable increase ($\sim$ 23 \% at $T_{FE}$ for 30 kOe) of $P$ is noted at higher temperature, pointing a significant magnetoelectric coupling. Synchrotron diffraction studies indicate that the structural changes are associated with the FE ordering, as observed from the [11$l$] scan. The amplitude refinement of the neutron diffraction results confirm that the shift of the apex O atoms of the FeO$_6$ octahedra along $c$-axis in the $Pbn2_1$ polar structure correlates the ferroelectricity. The change of slope in the thermal variation of [011] magnetic reflection around $T_{FE}$ proposes that the exchange striction mechanism leads to the ferroelectricity for HoFeO$_3$.

%%%%%%%%%%%%%%%%%%%%%%%%%%%%%%%%%%%%%%%%%%%%%%%%%%%%%%%%%%%%%%%%%%%%%%%%%%%%%%%%%%%%%%%%%%%%%%%%%%%%%%%%%%%%%%%%%%%%%%%
\begin{figure}[t]
\centering
\includegraphics[width = \columnwidth]{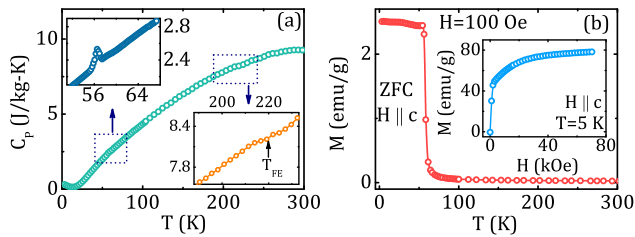}
\caption {Thermal variations of (a) heat capacity ($C_P$) and (b) magnetization ($M$) along the $c$-axis for HoFeO$_3$. Upper and lower insets of (a) magnify the low-$T$ and high-$T$ regions, respectively. Inset of (b) shows a magnetization curve at 5 K.} 
\label{C-mag}
\end{figure}
%%%%%%%%%%%%%%%%%%%%%%%%%%%%%%%%%%%%%%%%%%%%%%%%%%%%%%%%%%%%%%%%%%%%%%%%%%%%%%%%%%%%%%%%%%%%%%%%%%%%%%%%%%%%%%%%%%%%%%%
%
%

The HoFeO$_3$ single crystals were grown by S. N. Barilo and D. I. Zhigunov in Minsk, Belarus using flux method. Single crystal neutron diffraction study was performed on a neutron-size crystal, which was glued on a Vanadium pin and placed on the self-dedicated low-temperature Displex device \cite{arc}. Two full data sets were collected at 200 and 250 K at the D19 diffractometer (ILL, Grenoble) operating in high resolution mode with a wavelength of 0.94822 \AA. Datasets consist of omega-scans at selected $\chi$ and $\phi$ positions to obtain a completeness above 90\%. The data collection was performed using the MAD program. The reflection data were indexed with the ILL program PFIND and integrated with the RETREAT software\cite{mci} from the ILL program suite. The correction for attenuation was carried out with the ILL program D19ABS \cite{matt}. Neutron diffraction was further performed on a crystal with a 2.36 \AA~ wavelength in the D10 diffractometer of ILL. The crystal was mounted on the cold tip of the D10 cryostat with its [100] axis approximately parallel to the $\omega$-axis of the diffractometer. Single crystal synchrotron diffraction study was performed at beamline P09/PETRA III at DESY, Germany\cite{joerg}. The samples were mounted in a He-flow magnet cryostat covering the temperature range from 2 to 300 K with the cryostat mounted on a horizontal Psi-diffractometer. The experiment was conducted at photon energies significantly below the L2/3 absorption edges of the rare earth elements to avoid fluorescence.  Dielectric permittivity was recorded in a E4980A LCR meter (Agilent Technologies, USA) equipped with a PPMS-II system (Quantum Design, USA). The pyroelectric current was recorded using an electrometer (Keithley, model 6517B) at a constant temperature sweep rate. The $P-E$ loops were recorded using a FE loop tracer (Radiant Technology, USA). All electrical contacts were fabricated using an air drying silver paint. Heat capacity was measured in a PPMS-I system (Quantum Design, USA). The dc magnetization was measured in a commercial magnetometer of Quantum Design (MPMS, evercool).

%%%%%%%%%%%%%%%%%%%%%%%%%%%%%%%%%%%%%%%%%%%%%%%%%%%%%%%%%%%%%%%%%%%%%%%%%%%%%%%%%%%%%%%%%%%%%%%%%%%%%%%%%%%%%%%%%%%%%%%
\begin{figure}
\centering
\includegraphics[width = \columnwidth]{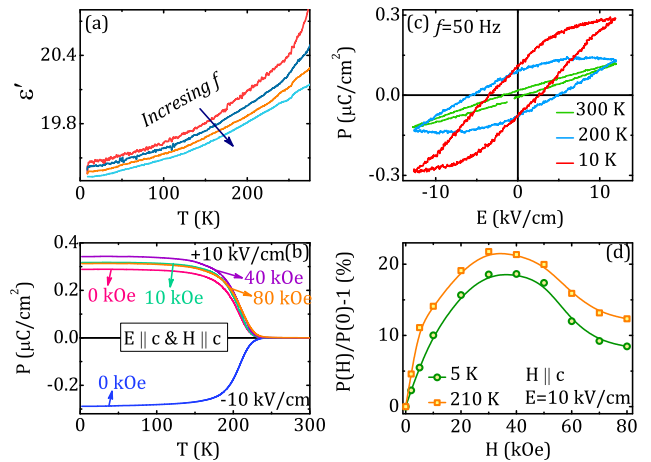}
\caption {Thermal variations of (a) real ($\epsilon^{\prime}$) part of dielectric permittivity at different $f$  = 2, 5, 10, and 20 kHz, (b) electric polarizations ($P$) for $E$ = $\pm$ 10 kV/cm and with selected $H$, (c) P$-$E loops at 10, 200, and 300 K. (d) Percentage of change of polarization with $H$ at 5 and 210 K.} 
\label{polar}
\end{figure}
%%%%%%%%%%%%%%%%%%%%%%%%%%%%%%%%%%%%%%%%%%%%%%%%%%%%%%%%%%%%%%%%%%%%%%%%%%%%%%%%%%%%%%%%%%%%%%%%%%%%%%%%%%%%%%%%%%%%%%%

Heat capacity ($C_P$) in zero-field with temperature ($T$) is depicted in Fig. \ref{C-mag}(a) for HoFeO$_3$. Upper and lower insets provide the low-$T$ and high-$T$ regions, highlighting the spin re-orientation transition at 56 K\cite{koe_PR,bujko_JPCM,bhatt_JPCS,shao_JCG,nikol_JPCM,chat_AIP} and an   anomaly close to  $T_{FE}$, respectively. A change of slope in $C_P(T)$ is observed around $T_{FE}$, which is correlated with the symmetry breaking, also confirmed by the synchrotron and neutron diffraction results. The sharp rise in ZFC magnetization ($M$) at 56 K points to a spin re-orientation transition, while any noticeable change is absent around $T_{FE}$, as depicted in Fig. \ref{C-mag}(b) for $H$ = 100 Oe along $c$-axis. A magnetization curve at 5 K is depicted in the inset of Fig. \ref{C-mag}(b). It shows an initial sharp rise with $H$, followed by a saturating trend above $\sim$ 40 kOe.

%%%%%%%%%%%%%%%%%%%%%%%%%%%%%%%%%%%%%%%%%%%%%%%%%%%%%%%%%%%%%%%%%%%%%%%%%%%%%%%%%%%%%%%%%%%%%%%%%%%%%%%%%%%%%%%%%%%%%%%
\begin{figure}[t]
\centering
\includegraphics[width = \columnwidth]{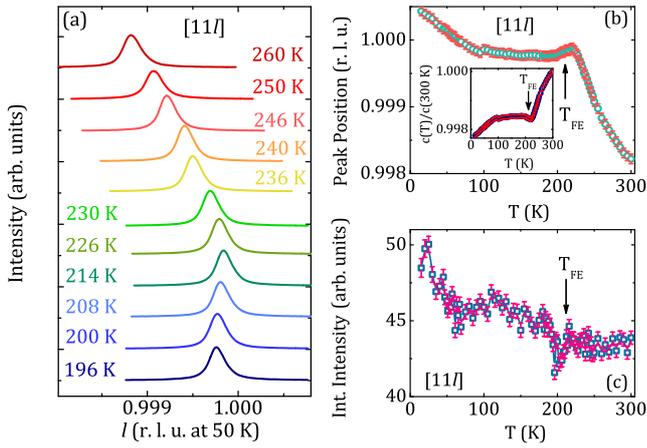}
\caption {(a) The [11$l$] scan at selected temperatures around $T_{FE}$. The $T$ variations of (b) peak positions and (c) integrated intensity of [11$l$] scan. Inset of (b) shows the variation of $c(T)/c(300$ K) with $T$.} 
\label{XRD}
\end{figure}
%%%%%%%%%%%%%%%%%%%%%%%%%%%%%%%%%%%%%%%%%%%%%%%%%%%%%%%%%%%%%%%%%%%%%%%%%%%%%%%%%%%%%%%%%%%%%%%%%%%%%%%%%%%%%%%%%%%%%%%

Dielectric permittivity ($\epsilon$) is measured along the $c$-axis with $T$ at different frequencies ($f$) in the range of 2 - 20 kHz. The real ($\epsilon^{\prime}$) components of $\epsilon$ are plotted with $T$ in Fig. \ref{polar}(a). Any convincing signature of $T_{FE}$ is absent. The absence of  signature in the high temperature may be attributed to the overlapping of the intrinsic component with the extrinsic components such as the grain boundary and the sample-electrode interface effects.\cite{ghosh_EPL,indra1,indra2} Appearance of a ferroelectric order is first detected along the $c$-axis from the pyroelectric current measurements, while the sample is cooled from different poling temperatures ($T_{pole}$), as shown in Fig. S1 of the supplemental information (SI) \cite{SI}. For all $T_{pole}$s, a peak around $\sim$ 210 K is always detected and confirms genuine occurrence of a polar order.   
Time integrated pyroelectric current measurements provide $P(T)$ for the $\pm$ 10 kV/cm poling field along the crystallographic $c$-axis, which is depicted in Fig. \ref{polar}(b). Our attempts fail to detect the ferroelectricity along the $a$- and $b$-axes. The result is similar to the observed ferroelectricity along the $c$-axis for GdFeO$_3$\cite{toku_NM} and DyFeO$_3$\cite{toku_PRL}. Reversal of $P$ for $E$ = -10 kV/cm corroborates the ferroelectricity. Figure \ref{polar}(c) depicts the $P-E$ loops at 10, 200, and 300 K. At 300 K the loop ascribes to the lossy current. The value of $P$ is $\sim$ 0.29 $\mu$C/cm$^2$ at 10 K, which is comparable to those reported for the promising orthoferrites\cite{toku_NM} and manganites\cite{kimu_NAT}. The value of $P$ decreases at 200 K in accordance with the $T$ variation of $P(T)$. As shown in Fig. \ref{polar}(b), the $P(T)$ is strongly influenced by $H$ along the $c$-axis, pointing to a significant ME coupling. The percentage of changes in $P$ defined as $\Delta P$(\%) = [$P(H)/P(0) - 1]\times$100 at 5 K and $T_{FE}$ are depicted as a function of $H$ in Fig. \ref{polar}(d). Here, $P(0)$ is recorded at zero field. $\Delta P$(\%) increases with $H$ and a maximum is reached for both temperatures in the field range of 30-40 kOe, above which it decreases till 80 kOe. We note that the changes in $\Delta P$(\%) at 5 K may be correlated with the magnetization curve at 5 K, as shown in the inset of Fig. \ref{C-mag}(b) \cite{ankita_new}. Here, the increase of $\Delta P$(\%) is associated with the increase of $M$ initially and the decrease of $\Delta P$(\%) is noted above $\sim$ 40 kOe, above which a saturating trend of $M$ is observed in the magnetization curve. The maximum value is achieved as high as $\sim$ 23 \%  at $T_{FE}$ for $H$ = 30 kOe.  

To investigate the structural correlation to the observed FE order, the synchrotron diffraction study\cite{joerg} is performed over a temperature range of 5$-$300 K. The changes in the peaks at selected $T$ are shown in Fig. \ref{XRD}(a). The peak positions and integrated intensities are depicted with $T$ in Figs. \ref{XRD}(b) and \ref{XRD}(c), respectively.  A maximum is observed in the peak position, which is slightly above $T_{FE}$. This points to a significant change in the lattice parameters around $T_{FE}$, because the change in $l$ is inversely proportional to the lattice constant, $c$. The inset of Fig. \ref{XRD}(b) depicts the $T$ variation of $c(T)$, scaled by the value at 300 K [$c(T)/c(300$ K)]. An anomalous thermal expansion of $c(T)/c(300$ K) is observed around $T_{FE}$. With decreasing $T$ from 300 K, the integrated intensity remains unchanged till $T_{FE}$, below which it decreases, exhibiting a 'dip'. Below $\sim$ 200 K, it increases down to the low temperature, exhibiting an another 'dip' close to $\sim$ 55 K. This is close to the spin reorientation transition, pointing a magnetoelastic coupling. Here, the change in the integrated intensities around $T_{FE}$ is significant, because it involves a change in the scattering cross section and may point to a possible structural transition, as adequately reported for the iso-structural transition\cite{dey1_PRB,dey2_PRB,dey3_PRB} as well as symmetry lowering\cite{hori_PRL,koo_PRL,ghosh_EPL,indra_JPCM} around the FE ordering. 

%
%%%%%%%%%%%%%%%%%%%%%%%%%%%%%%%%%%%%%%%%%%%%%%%%%%%%%%%%%%%%%%%%%%%%%%%%%%%%%%%%%%%%%%%%%%%%%%%%%%%%%%%%%%%%%%%%%%%%%%%
\begin{figure}[t]
\centering
\includegraphics[width = \columnwidth]{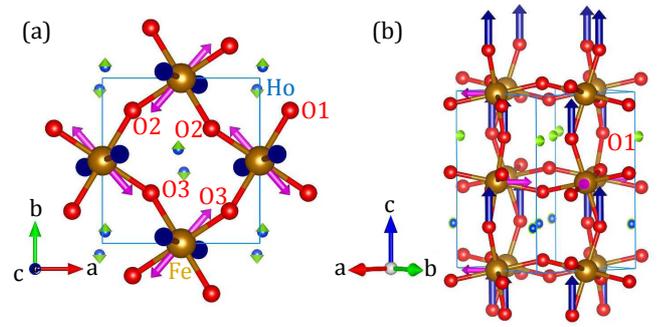}
\caption {The atomic displacements are depicted in different directions. (a) The directions of small Ho (green) and considerable Fe (pink) displacements for the projection along $c$-axis. (b) The displacement of O1 atoms (deep blue) along $c$-axis is shown. A fix scale of 10 is used for all the active modes.} 
\label{ND}
\end{figure}
%%%%%%%%%%%%%%%%%%%%%%%%%%%%%%%%%%%%%%%%%%%%%%%%%%%%%%%%%%%%%%%%%%%%%%%%%%%%%%%%%%%%%%%%%%%%%%%%%%%%%%%%%%%%%%%%%%%%%%%

We note that a $G$-type ${\bf k}$ = 0 AFM structure has been proposed\cite{koe_PR,chat_AIP,chat_arxiv}, around which ferroelectricity occurs in HoFeO$_3$. Thus, the inverse Dzyaloshinskii-Moriya (DM) interaction does not contribute to the observed ferroelectricity. To resolve this, a neutron diffraction study is performed on a HoFeO$_3$ crystal at two temperatures, above (250 K) and below (200 K) $T_{FE}$. Refinements are performed by taking into account of both, the displacive and magnetic symmetry modes using a single crystal Amplitude Fullprof program, where the pcr file has been created using the ISODISTORT suit\cite{isodistort}. At 250 K the $Pbnm$ space group is used as a parent space group. At 200 K the refinements are performed using both, $Pbnm$ and $Pbn2_1$, as the parent space groups, where the polar $Pbn2_1$ is a subgroup of $Pbnm$. We note that the R-factor improves from 3.96 to 3.76 for the refinement using $Pbn2_1$ as a parent space group at 200 K, which has been considered in the reported neutron results.\cite{chat_AIP,chat_arxiv} Although the symmetry is lowered, here, the final refinement is stable and converges to reasonable values. More importantly, it explains the ferroelectricity.

%%%%%%%%%%%%%%%%%%%%%%%%%%%%%%%%%%%%%%%%%%%%%%%%%%%%%%%%%%%%%%%%%%%%%%%%%%%%%%%%%%%%%%%%%%%
\begin{figure}[t]
\centering
\includegraphics[width = \columnwidth]{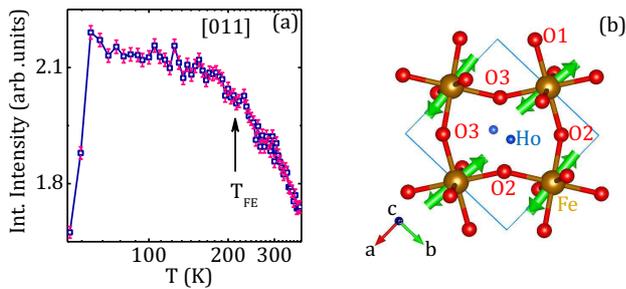}
\caption{$T$ variations of integrated intensity of the magnetic [011] reflections. (b) Proposed magnetic structure projected in the $ab$ plane.}
\label{ND_mag}
\end{figure}
%%%%%%%%%%%%%%%%%%%%%%%%%%%%%%%%%%%%%%%%%%%%%%%%%%%%%%%%%%%%%%%%%%%%%%%%%%%%%%%%%%%%%%%%%%%%

Out of four different Shubnikov space groups, such as $Pbn2_1$, $Pbn^{\prime}2_1^{\prime}$, $Pb^{\prime}n2_1^{\prime}$ and $Pb^{\prime}n^{\prime}2_1$, the last one only reproduces the magnetic structure at 200 K\cite{chat_AIP,chat_arxiv}. Only two irreducible representations are needed to distort the $Pbn2_1$ parent space group into the $Pb^{\prime}n^{\prime}2_1$ Shubnikov space group. Here, we discuss only GM1+, because it is the only displace mode in our system. In this Shubnikov group there are 15 possible modes vectors acting on the nuclear part at 200 K, where only four modes such as, A1, A4, A5, and A9 are active for the refinement using $Pbn2_1$, as a parent space group. The details are further discussed in SI \cite{SI}. The amplitude A1, operating on the Ho(III) atoms with an amplitude of 0.0021(6) \AA, gives rise to a weak shift in the $ab$-plane. However, this amplitude can not be responsible for the global ferroelectricity, because the polarization due to the displacement of a Ho atom is compensated by the same displacement with an opposite direction by the other Ho atom inside the unit cell, as more clearly depicted by the green small arrows in Fig. \ref{ND}(a). A similar behavior is observed for the Fe atoms. In this case two different amplitudes, A4 and A5 with amplitudes of -0.011(3) and -0.009(2) \AA, are present. The combination of these two modes also produces a shift of the iron atoms in the $ab$-plane close to the Fe-O2 and Fe-O3 bonds.\cite{O2} However, as in the case of Ho, these vectors are strictly compensated within the unit cell, as shown by the pink arrows in Fig. \ref{ND}(a). Therefore, the Fe displacement does not contribute to the global polarization. Out of four amplitudes, the last amplitude, A9, of -0.022(5) \AA acts on O1 of the FeO$_6$ octahedra, producing a global shift along the crystallographic $c$-axis. This displacement has the same direction for all the O1 within the unit cell, as clearly depicted by the deep blue arrows in Fig. \ref{ND}(b) and is accountable for the macroscopic electric polarization. This result is consistent with the anomalous expansion in $c(T)/c(300$ K) near $T_{FE}$. Using the atomic displacement of O1, the value of $P$ is $\sim$ 1.2 $\mu C$/cm$^2$ using a simplified formula, $P = 1/V [\sum_i(m_i\Delta z_{i}Z_i)e$], where $m_i$ is the crystallographic site multiplicity, $\Delta z_i$ is the displacement along the polar axis (here, crystallographic $c$-axis), $Z_{i}e$ is the charge, and $V$ is the unit cell volume. The calculated value is $\sim$ 3 times higher than the value of $P$, as obtained for the poling field of 10 kV/cm. Instead of point charge model, the Berry phase method using a first-principles calculation would provide a more realistic value of $P$ \cite{berry}. We note that the proposed polar $Pbn2_1$ structure has also been proposed for few members of the $R$CrO$_3$ series \cite{ghosh_EPL,indra_JPCM,mahana} and SmFeO$_3$\cite{zhang_JALCOM} for accounting ferroelectricity. 

To understand the possible origin of the ferroelectric order, magnetic reflections are recorded as a function of temperature. The variations of the integrated intensities of the magnetic reflections along selected directions are shown in Fig. S5 of SI \cite{SI}. Out of all the recorded magnetic reflections, the [011] reflection is the most sensitive for the proposed magnetic structure, as shown in Fig. S4 of SI \cite{SI}. The temperature variation of the [011] reflection is depicted in Fig. \ref{ND_mag}(a), pointing to a change of slope around $T_{FE}$. The sharp fall of intensity around $\sim$ 55 K further confirms the spin reorientation transition, as also evident in the magnetization and heat capacity results.\cite{chat_AIP} The proposed magnetic structure at 200 K, described on $Pb^{\prime}n^{\prime}2_1$, is shown in Fig. \ref{ND_mag}(b), which is projected in the $ab$ plane. As obtained from the refinement, only one component of the three magnetic amplitudes is dominant and the value is 3.72(3) $\mu_B$ along $a$-axis at 200 K. The other two components at 200 K are 0.39(3) and 0.274(2) $\mu_B$ along the $b$- and $c$-axes, respectively. Each magnetic moment is not only antiferromagnetically coupled with the neighbors within the $ab$-plane but also along the $c$-axis, giving rise to a global antiferromagnetic behavior, as depicted in Fig. S4 of SI \cite{SI}. We note that above $T_{FE}$ a similar refinement at 250 K  provides the magnetic amplitudes, as 3.79(3), 0.39(3), and 0.3(2) $\mu_B$ along the $a$-, $b$-, and $c$-axes, respectively, where $Pb^{\prime}n^{\prime}m$ is used for the refinement. The slight changes in magnetic amplitudes below and above $T_{FE}$ are consistent with the slope change in the [011] reflection intensity around $T_{FE}$. The results correlate with the structural change and indicate that the exchange striction mechanism causes the ferroelectricity in HoFeO$_3$. We note that the exchange striction mechanism driven ferroelectricity has been discussed by Khomskii for addressing type-II multiferroics.\cite{khom} 

In conclusion, ferroelectric order is observed near ambient temperature with $T_{FE}$ at 210 K for HoFeO$_3$. The reasonably high value of the electric polarization ($\sim$ 0.29 $\mu$C/cm$^2$) associated with a strong magnetoelectric coupling at low magnetic field attracts the community. Synchrotron and neutron diffraction studies of the single crystal propose that the breaking of space inversion symmetry in polar $Pbn2_1$ structure correlates with the ferroelectricity. More specifically, the apex oxygen displacement of the FeO$_6$ octahedra engineers the ferroelectric polarization along the crystallographic $c$-axis. A possible exchange striction mechanism is suggested for interpreting the ferroelectricity in HoFeO$_3$. This work contributes a significant progress in searching new type-II multiferroics with a high-temperature magnetoelectric coupling.

\vspace{0.2in}
\noindent
{\bf Acknowledgment}\\
The authors would like to thank J. Rodr\'{i}guez-Carvajal for discussions and integrating the Amplitude refinement program to the Fullprof suite. S.G. acknowledges the financial support from DST, India (Project No. SB/S2/CMP-029/2014), from BRNS, India for the neutron diffraction studies at ILL, France, and from DST-DESY project (No. I-20140424) for the synchrotron diffraction experiment at DESY, Germany.

\end{document}